\documentclass[12pt]{article}
\usepackage{graphicx}
\makeatletter
	
		\@addtoreset{equation}{section}
\makeatother

\newcommand{\mapright}[2]
{\mathop{\hbox to 8mm{\rightarrowfill}}
\limits^{\scriptstyle #1}_{\scriptstyle #2}}

\makeatletter
\@addtoreset{equation}{section}
\makeatother

\begin{document} %
\begin{flushright}
{KOBE-TH-03-08}\\
{OU-HET-458}\\
\end{flushright}
\vspace{5mm}
\vspace*{6mm}
\begin{center}
{\Large \bf
Supersymmetry and discrete transformations \\ on $S^1$ 
with point singularities}
\vspace{10mm} \\
Tomoaki Nagasawa\footnote{e-mail:
nagasawa@phys.sci.kobe-u.ac.jp (T. Nagasawa)},
Makoto Sakamoto\footnote{e-mail:
sakamoto@phys.sci.kobe-u.ac.jp (M. Sakamoto)},
Kazunori Takenaga\footnote{e-mail:
takenaga@het.phys.sci.osaka-u.ac.jp (K. Takenaga)}
\vspace*{10mm} \\
{\small \it
${}^1 $Graduate School of Science and Technology, Kobe University,
Rokkodai, Nada, \\ Kobe 657-8501, Japan
\\
${}^2 $Department of Physics, Kobe University,
Rokkodai, Nada, Kobe 657-8501, Japan
\\
${}^3 $Department of Physics, Osaka University, 
Toyonaka 560-0043, Japan
}
\vspace{15mm}
\\
{\bf Abstract}
\vspace{4mm}
\\
\begin{minipage}[t]{130mm}
\baselineskip 6mm 

We investigate $N$-extended supersymmetry in one-dimensional 
quantum mechanics on a circle with point singularities. 
For any integer $n$, $N=2n$ supercharges are explicitly 
constructed and a class of point singularities compatible 
with supersymmetry is clarified.
Key ingredients in our construction are $n$ sets of 
discrete transformations, each of which forms 
an $su(2)$ algebra of spin $1/2$.
The degeneracy of the spectrum and spontaneous supersymmetry 
breaking are briefly discussed.
\end{minipage}
\end{center}
\vspace{8mm}
\noindent
%
\newpage
\baselineskip 6mm 

\section{Introduction}

Quantum mechanics in one dimension admits point singularities 
as interactions of zero range.
A point singularity is parameterized by the group 
$U(2)$ \cite{u(2)1, u(2)2,  u(2)3}, and the parameters characterize 
connection conditions between a wavefunction and its derivative 
at the singularity.
The variety of the connection conditions
leads to various interesting physical phenomena, 
such as duality \cite{duality1,duality2}, the Berry phase 
\cite{berry1, berry2}, scale anomaly \cite{anomaly} and 
supersymmetry \cite{susy1,susy2,susy3,susy5}.
Since a system with a number of point singularities possesses 
a wider parameter space, the system can have new features.
In this Letter, we show that $2^n$ point singularities 
on a circle can realize $N=2n$ supersymmetry.

In Ref. \cite{susy1}, $N=2$ supersymmetry was discussed 
in the system of a free particle on a line ${\bf R}$ or 
an interval $[-l,l]$ with a point singularity.
In Ref. \cite{susy3}, this work was extended to $N=4$ supersymmetry 
in the system on a pair of lines ${\bf R}$ or intervals $[-l,l]$ 
each having a point singularity.
In Ref. \cite{susy2}, two point singularities were put 
on a circle, and $N=2$ supersymmetric models with a superpotential 
were constructed.
Spectral properties and domains of operators 
in a supersymmetry algebra
on a circle with a singularity are discussed 
in Ref. \cite{susy5}.

The purpose of this Letter is to examine $N$-extended supersymmetry 
in quantum mechanics on a circle with point singularities.
Since it is hard to extend the work of Ref. \cite{susy3} 
to higher $N$-extended supersymmetry, we follow the approach of 
Ref. \cite{susy2} to realize $N=2n$ supersymmetry for any integer $n$.
A key ingredient in the analysis of Ref. \cite{susy2} is 
a set of the discrete transformations that forms an $su(2)$ 
algebra of spin $1/2$.
In the next section, we extend it to $n$ sets of discrete 
transformations, each of which forms an $su(2)$ algebra of spin $1/2$
and commutes with the others.
Since these transformations, in general, make wavefunctions 
discontinuous at some points on a circle, point singularities 
are inevitable in our construction of $N$-extended supersymmetry.

In Section 2, we introduce $n$ sets of discrete transformations, 
each of which forms an $su(2)$ algebra of spin $1/2$.
In Section 3, we construct $N=2n$ supercharges in terms of 
these transformations.
In Section 4, we find a class of connection conditions compatible 
with supersymmetry. 
In Section 5, the degeneracy of the spectrum and spontaneous 
supersymmetry breaking are briefly discussed. 
Section 6 is devoted to summary  and discussions.
\section{Discrete transformations}

The system we consider is one-dimensional quantum mechanics 
on a circle $S^1 (-l< x \le l)$ with $2^n$ point singularities 
placed at 
\begin{equation}
	x= l_s \equiv \left(1-\frac{s}{2^{n-1}}\right) l, \qquad  
	{\rm for}\  s=0,1,\cdots, 2^{n}-1.
\end{equation}
In this setup, we are allowed to introduce $n$ sets of discrete 
transformations $\{ {\cal P}_k, {\cal Q}_k, {\cal R}_k \}$
$(k=1,2,\cdots, n)$ which produce singularities at $x=l_s$ 
for $s=0,1,\cdots,2^n -1$.
The ${\cal P}_k$ $(k=1,2,\cdots,n)$ are a kind of the parity 
transformation, and the action of ${\cal P}_k$ on an arbitrary 
function $\varphi(x)$ is defined by
\begin{eqnarray}
	({\cal P}_k \varphi) (x) 
		&=&
		 \sum_{s=1}^{2^{k-1}} 
		\Theta \left( x- \left(1-\frac{s}{2^{k-2}}\right)l\right)
		\Theta \left( \left( 1-\frac{s-1}{2^{k-2}}\right)l-x\right) 
		\nonumber \\
	&& \quad \times 
		\varphi \left( -x+\left(2-\frac{2s-1}{2^{k-2}}\right)l\right), 
	    \qquad {\rm for} \  k=1,2,\cdots,n,
\end{eqnarray}
where $\Theta(x)$ is the Heaviside step function. 
The ${\cal P}_1$ is just a familiar parity transformation, 
$({\cal P}_1 \varphi)(x) =\varphi(-x)$.
An example of the action of ${\cal P}_3$ on a function $\varphi(x)$ 
is given in Fig. 1.
We see that the action of ${\cal P}_3$, in general, 
produces singularities at $x=0,\pm\frac{l}{2}$ and $l$.
The action of ${\cal P}_k$ for $k=1,2$ and $3$ is schematically 
depicted in Fig. 2.
The ${\cal R}_k$ $(k=1,2,\cdots,n)$ are a kind of the half-reflection 
transformation, and the action of ${\cal R}_k$ on $\varphi(x)$ is 
defined by
\begin{eqnarray}
	({\cal R}_k \varphi)(x) 
	&=& \sum_{s=1}^{2^{k-1}} (-1)^s 
		\left[ -\Theta \left(x-\left( 1-\frac{s-1/2}{2^{k-2}}\right)l\right)
			\Theta\left( \left( 1-\frac{s-1}{2^{k-2}}\right)l-x\right) 
			\right.\nonumber \\
	&& \qquad \qquad 	\quad 
		\left. +\Theta \left( x- \left( 1- \frac{s}{2^{k-2}}\right) l \right) 
		\Theta\left(\left( 1-\frac{s-1/2}{2^{k-2}}\right)l -x \right) \right]
		\varphi(x), \nonumber \\
	&& \qquad \qquad \qquad  \qquad \qquad {\rm for} \ k=1,2,\cdots,n.
\end{eqnarray}
For $k=1,2$ and $3$, $({\cal R}_k \varphi )(x)$ are explicitly given by 
\begin{eqnarray}
	&&({\cal R}_1 \varphi )(x) =
		\left\{
			\begin{array}{l}
				+ \varphi(x) \quad {\rm for}\ \;    0<x<l, \\
				- \varphi(x) \quad {\rm for}\   -l<x<0,
			\end{array}
		\right.
	\\
	&&({\cal R}_2 \varphi )(x) =
		\left\{
			\begin{array}{l}
				+ \varphi(x) \quad {\rm for} \ -l<x<-\frac{l}{2} 
				\  {\rm and} \ \frac{l}{2}<x<l, \\
				- \varphi(x) \quad {\rm for} \ -\frac{l}{2}<x<\frac{l}{2},
			\end{array}
		\right.
	\\
	&&({\cal R}_3 \varphi )(x) =
		\left\{
			\begin{array}{lc}
				+ \varphi(x) \quad {\rm for} \  -l<x<-\frac{3l}{4}, 
				-\frac{l}{4}<x<\frac{l}{4} \ {\rm and} \  \frac{3l}{4}<x<l, \\
				- \varphi(x) \quad {\rm for} \ -\frac{3l}{4}<x<-\frac{l}{4} 
				\ {\rm and}\  \frac{l}{4}<x<\frac{3l}{4},
			\end{array}
		\right.
\end{eqnarray}
and are schematically depicted in Fig. 3.
The third transformations ${\cal Q}_k$ $(k=1,2,\cdots,n)$ are defined, 
in terms of ${\cal P}_k$ and ${\cal R}_k$, by
\begin{equation}
	{\cal Q}_k \equiv -i {\cal R}_k {\cal P}_k \quad {\rm for} 
	\  k=1,2,\cdots,n.
\end{equation}

Important observations are that each set of $\{ {\cal P}_k, 
{\cal Q}_k, {\cal R}_k \}$ $(k=1,2,\cdots,n)$ forms an $su(2)$ 
algebra of spin $1/2$, i.e.,
\begin{eqnarray}
	&&{\cal P}_k {\cal Q}_k =- {\cal Q}_k {\cal P}_k =i {\cal R}_k, 
	\nonumber \\
	&&{\cal Q}_k {\cal R}_k =- {\cal R}_k {\cal Q}_k =i {\cal P}_k, 
	\nonumber \\
	&&{\cal R}_k {\cal P}_k =- {\cal P}_k {\cal R}_k =i {\cal Q}_k, 
	\nonumber \\
	&& ({\cal P}_k )^2 =({\cal Q}_k)^2 =({\cal R}_k)^2 =1,
\end{eqnarray}
and 
\begin{equation}
	 [ {\cal O}_k , {\cal O}_{k'} ]=0 \quad {\rm if} \  k \ne k'
\end{equation}
for ${\cal O}_k \in \{ {\cal P}_k, {\cal Q}_k, {\cal R}_k \}$ 
and ${\cal O}_{k'}=\{ {\cal P}_{k'}, {\cal Q}_{k'}, {\cal R}_{k'} \}$.

For later use, let us introduce new sets of $su(2)$ generators 
$\{ {\cal G}_{{\cal P}_k}, {\cal G}_{{\cal Q}_k}, 
{\cal G}_{{\cal R}_k} \}$ $(k=1,2,\cdots,n)$ as
\begin{equation}
	{\cal G}_{\vec{{\cal P}}_k} ={\cal V}_k^{\dagger} 
	\vec{{\cal {P}}}_k {\cal V}_k, \quad {\rm for} \  k=1,2,\cdots,n
\label{s-u-t}
\end{equation}
for ${\cal V}_k = e^{i \vec{v}_k \cdot 
\vec{{\cal P}}_k}\in SU(2)$.
Here, we have used $\vec{{\cal P}}_k$ as an abbreviation of 
$\vec{{\cal P}}_k = ({\cal P}_k, {\cal Q}_k, {\cal R}_k).$
The new $su(2)$ generators ${\cal G}_{\vec{{\cal P}}_k}$ have to be 
linearly related to $\vec{{\cal P}_k}$ as 
\begin{eqnarray}
	{\cal G}_{{\cal P}_k} &=& \vec{e}_{{\cal P}_k}\cdot \vec{{\cal P}_k} , 
	\nonumber \\
	{\cal G}_{{\cal Q}_k} &=& \vec{e}_{{\cal Q}_k}\cdot \vec{{\cal P}_k} , 
	\nonumber \\
	{\cal G}_{{\cal R}_k} &=& \vec{e}_{{\cal R}_k}\cdot \vec{{\cal P}_k} , 
	\quad {\rm for} \ k=1,2,\cdots,n,
\end{eqnarray}
where $\{ \vec{e}_{{\cal P}_k}, \vec{e}_{{\cal Q}_k}, 
\vec{e}_{{\cal R}_k}\}$ are three-dimensional orthogonal unit vectors.
One might think that the transformation (\ref{s-u-t}) is merely 
a change of the basis of the $su(2)$ generators and does not change 
physics.
This is not, however, the case.
It should be emphasized that the transformation (\ref{s-u-t}) is 
a {\it singular} unitary transformation because it, in general, 
changes connection conditions of wavefunctions at singular points.
The transformation may be regarded as a duality connecting 
different theories (with different connection conditions).
\section{$N=2n$ superalgebra}

In Ref. \cite{susy2}, $N=2$ supercharges are constructed in the 
system on a circle with two point singularities placed at $x=0$ and $l$.
An extension of the supercharges to $N=2n$ supercharges for 
any integer $n$ will be given by
\begin{equation}
	Q_a =\frac{i}{2} \Gamma_a {\cal D}, \quad {\rm for} \  
	a=1,2,\cdots,2n,
	\label{2ns-charges}
\end{equation}
where
\begin{eqnarray}	
	 {\cal D} &=& {\cal R}_1 \cdots {\cal R}_n \frac{d}{dx} + 
	    {\cal G}_{{\cal R}_1}\cdots {\cal G}_{{\cal R}_n}
	    {\cal R}_1\cdots{\cal R}_n W'(x),  \\
	 \Gamma_{2k-1} &=& {\cal G}_{{\cal R}_1}\cdots 
	    {\cal G}_{{\cal R}_{k-1}}{\cal G}_{{\cal P}_k},  \\
	 \Gamma_{2k} &=& {\cal G}_{{\cal R}_1}\cdots 
	    {\cal G}_{{\cal R}_{k-1}}{\cal G}_{{\cal Q}_k}, 
	    \quad {\rm for} \  k=1,2,\cdots,n.
\end{eqnarray}
Here, $W'(x)=\frac{d}{dx}W(x)$ and $W(x)$ is called a superpotential.
The function $W'(x)$ is allowed to have discontinuities 
at singular points $x=l_s$ $(s=0,1,\cdots,2^n-1)$ and is 
assumed to obey
\begin{equation}
	{\cal P}_k W'(x) = -W'(x) {\cal P}_k , \quad {\rm for} 
	\  k=1,2,\cdots,n.
\end{equation}
Noting that ${\cal R}_1\cdots {\cal R}_n \frac{d}{dx}$ and 
${\cal R}_1\cdots {\cal R}_n W'(x)$ commute with $\vec{{\cal P}_k}$ 
for $^{\forall} k=1,2,\cdots,n$, we can show that the supercharges 
$Q_a$ $(a=1,2,\cdots,2n)$ form the $N=2n$ superalgebra
\begin{equation}
	\{ Q_a, Q_b \} =H \delta_{ab} , \quad {\rm for} \ a,b=1,2,\cdots,2n
	\label{super-algebra}
\end{equation}
with the Hamiltonian 
\begin{equation}
	H=\frac{1}{2} \left[
		-\frac{d^2}{dx^2} - {\cal G}_{{\cal R}_1}\cdots 
		{\cal G}_{{\cal R}_n} W''(x)+ \left(W'(x)\right)^2
	\right].
	\label{hamiltonian}
\end{equation}
\section{Compatibility with supersymmetry}

It is important to realize that our quantum system is 
specified by not only the Hamiltonian but connection conditions 
for wavefunctions.
This is because the system contains point singularities and
we need to impose 
appropriate connection conditions there. 
The Hilbert space is then defined by a space spanned
by eigenfunctions of the Hamiltonian (\ref{hamiltonian})
with the connection conditions
which have to make the Hamiltonian hermitian.
This setting is not, however, enough to guarantee the
$N=2n$ supersymmetry of the theory,
because the hermiticity of the Hamiltonian
does not, in general, assure the hermiticity of the supercharges 
$Q_a$ $(a=1,2,\cdots,2n)$ 
and further because for any state $\varphi(x)$ of the
Hilbert space $Q_{a}\varphi(x)$ 
do not, in general, belong to the same Hilbert space
(i.e., $Q_{a}\varphi(x)$ do not, in general, obey the same
connection conditions as $\varphi(x)$).
The supercharges would be then ill defined on the Hilbert space.

To give allowed connection conditions compatible with the 
$N=2n$ supersymmetry, 
let us introduce a $2^{n+1}$-dimensional 
vector $\Phi_{\varphi}$ that consists of boundary values of 
a wavefunction $\varphi(x)$ at the singularities, i.e., 
$\varphi(l_s \pm \varepsilon)$ for $s=0,1,\cdots,2^{n}-1$ 
with an infinitesimal positive constant $\varepsilon$.
It is convenient to arrange $\varphi(l_s \pm \varepsilon)$ 
in such a way that $\Phi_{\varphi}$ satisfies the relations
\begin{eqnarray}
	&&\Phi_{{\cal P}_k \varphi} = 
		(\overbrace{I_2 \otimes \cdots \otimes I_2 \otimes 
		\sigma_1}^{k} \otimes I_2 \otimes \cdots \otimes I_2) 
		\Phi_{\varphi}, \\
	&&\Phi_{{\cal Q}_k \varphi} = 
		(I_2 \otimes \cdots \otimes I_2 \otimes \sigma_2 
		\otimes I_2 \otimes \cdots \otimes I_2) \Phi_{\varphi}, \\
	&&\Phi_{{\cal R}_k \varphi} = 
		(\underbrace{ I_2 \otimes \cdots \otimes I_2 \otimes 
		\sigma_3\otimes I_2 \otimes \cdots \otimes I_2}_{n+1}) 
		\Phi_{\varphi},
\end{eqnarray}
where $I_M$ denotes an $M \times M$ unit matrix. 
For instance, $\Phi_{\varphi}$ for $n=1$ will be given by
\begin{equation}
	\Phi_{\varphi} = \left(\varphi(l-\varepsilon), 
	\varphi(0+\varepsilon), \varphi(-l+\varepsilon), 
	\varphi(0-\varepsilon) \right)^{T}.
\end{equation}
For $n=2$, $\Phi_{\varphi}$ will be given by 
\begin{eqnarray}
	\Phi_{\varphi} &=& \left( 
		\varphi(l-\varepsilon), \varphi\left(l/2+
		\varepsilon\right), \varphi(0+\varepsilon),
		\varphi\left(l/2-\varepsilon\right), 
		\right. \nonumber \\
	&& \quad 
		\left.
			\varphi(-l+\varepsilon), \varphi
			\left(-l/2-\varepsilon\right), 
			\varphi(0-\varepsilon), \varphi\left(-l/2+
			\varepsilon\right) \right)^T .
\end{eqnarray}

Let us consider the following type of connection conditions for
a wavefunction $\varphi(x)$:
\begin{eqnarray}
	&&\left(I_{2^{n+1}}-U \right)\Phi_{\varphi}=0,  \label{cc1}\\
	&&\left(I_{2^{n+1}}+U\right)\Sigma_3 \Phi_{{\cal D}\varphi}=0,  
	\label{cc2}
\end{eqnarray}
where $U$ is any $2^{n+1}\times2^{n+1}$ matrix satisfying
\begin{eqnarray}
	\Sigma_3 \gamma_a U &=&-U\Sigma_3 \gamma_a , \quad {\rm for} 
	\  a=1,2,\cdots,2n, \\
	U^{\dagger}U &=&I_{2^{n+1}}, \\
	U^2&=&I_{2^{n+1}},\label{u2=1}
\end{eqnarray}
with
\begin{eqnarray}
	\Sigma_3 &=&I_2 \otimes \cdots \otimes I_2\otimes \sigma_3, \\
	\gamma_{2k-1}&=& \vec{e}_{{\cal R}_1}\cdot \vec{\sigma}\otimes 
	\cdots \otimes \vec{e}_{{\cal R}_{k-1}}\cdot \vec{\sigma}\otimes 
	\vec{e}_{{\cal P}_k}\cdot \vec{\sigma}\otimes I_2\otimes \cdots 
	\otimes I_2, \\
	\gamma_{2k}&= &\vec{e}_{{\cal R}_1}\cdot \vec{\sigma}\otimes \cdots 
	\otimes \vec{e}_{{\cal R}_{k-1}}\cdot \vec{\sigma}\otimes 
	\vec{e}_{{\cal Q}_k}\cdot \vec{\sigma}\otimes I_2\otimes \cdots 
	\otimes I_2, \\
	&&  \qquad \qquad \qquad \qquad \qquad \qquad {\rm for} \  
	k=1,2,\cdots,n. \nonumber 
\end{eqnarray}
The Hilbert space is then assumed to be spanned by eigenfunctions
of the Hamiltonian (\ref{hamiltonian}) satisfying the connection
conditions (\ref{cc1}) and (\ref{cc2}).
The last condition (\ref{u2=1}) implies that any eigenvalue of 
$U$ is $+1$ or $-1$, so that the total number of the constraints 
in Eqs. (\ref{cc1}) and (\ref{cc2}) is $2\times 2^n$. 
This is the correct number to solve the Schr{\"o}dinger equation 
in our system because two connection conditions between $\varphi(x)$ 
and $\varphi'(x)$ should be imposed at each point singularity 
and there are $2^n$ point singularities in the present model.

The hermiticity conditions of the supercharges
\begin{eqnarray}
	\int_{-l}^l dx \psi^* (x) (Q_a \varphi)(x) = 
	\int_{-l}^l dx (Q_a \psi )^* (x) \varphi(x),\quad 
	{\rm for} \ a=1,2,\cdots,2n
\end{eqnarray}
give the nontrivial constraints on boundary values of the wavefunctions 
\begin{equation}
	\Phi_{\psi}^{\dagger} \Sigma_3 \gamma_a \Phi_{\varphi}=0, 
	\quad {\rm for} \  a=1,2,\cdots,2n.
	\label{Q-hermite}
\end{equation}
To derive it, we may use the formula of integration by parts
\begin{equation}
	\int_{-l}^l dx \xi^*(x) \left(\frac{d}{dx}\eta(x)\right)
	=-\int_{-l}^l dx \left( \frac{d}{dx}\xi(x)\right)^* \eta(x) 
	 +\Phi_{\xi}^{\dagger}(\sigma_3 \otimes \cdots \otimes 
	 \sigma_3)\Phi_{\eta},
\end{equation}
where the functions $\xi(x)$ and $\eta(x)$ are assumed to be 
continuous everywhere except for point singularities.
It is easy to show that the conditions (\ref{Q-hermite}) are satisfied
if $\varphi(x)$ and $\psi(x)$ obey the connection conditions
(\ref{cc1}) and (\ref{cc2}).
We can further show that 
for any eigenfunction $\varphi(x)$ of the Hilbert space
$Q_{a}\varphi(x)$ $(a=1,2,\cdots, n)$
also obey the connection conditions (\ref{cc1}) and (\ref{cc2}).
This implies that for any state $\varphi(x)$ of the
Hilbert space any products of $Q_a$'s on $\varphi(x)$,
$Q_{a_1}Q_{a_2}\cdots Q_{a_m}\varphi(x)$, 
belong to the same Hilbert space as $\varphi(x)$.
It follows from the algebra (\ref{super-algebra}) that 
the Hamiltonian is hermitian, as it should be.
Therefore, the action of the supercharges on the Hilbert space
is well-defined, and the algebra (\ref{super-algebra}) guarantees
the $N=2n$ supersymmetry of the theory.

It turns out that the following two types of the matrix $U$ 
satisfy the desired relations:
\begin{itemize}
\item[(I)] Type I
\end{itemize}
\begin{equation}
	U_{\rm I} (\pm) =\pm \left( \vec{e}_{{\cal R}_1} \cdot 
	\vec{\sigma}\otimes \cdots \otimes \vec{e}_{{\cal R}_n}
	\cdot \vec{\sigma}\otimes  I_2 \right),\label{typeI}
\end{equation}
\begin{itemize}
\item[(II)] Type II
\end{itemize}
\begin{eqnarray}
	U_{\rm II}(a) &=& a_1 \left( I_2 \otimes \cdots 
	 \otimes I_2 \otimes \sigma_1 \right)
	  +a_2 \left( I_2 \otimes \cdots \otimes I_2 
    	\otimes \sigma_2 \right) \nonumber \\
	&&+a_3 \left( \vec{e}_{{\cal R}_1}\cdot \vec{\sigma} 
    	\otimes \cdots \otimes \vec{e}_{{\cal R}_n}\cdot \vec{\sigma} 
	      \otimes \sigma_3 \right) \label{typeII}
\end{eqnarray}
with $a_1,a_2,a_3 \in $ {\bf R} and $(a_1)^2+(a_2)^2+(a_3)^2=1$.
The connection conditions found in Ref. \cite{susy2} correspond 
to the type I and the type II solutions with $a_1=a_2=0$ for $n=1$.
Although the configuration spaces in Ref. \cite{susy1,susy3} 
are different from ours, the results seem to be consistent 
with ours for $n=1$ and $n=2$ with a free Hamiltonian.
In our derivation, it is unclear whether the solutions 
(\ref{typeI}) and (\ref{typeII}) exhaust all allowed 
connection conditions compatible with supersymmetry.
This issue will be discussed in a forthcoming paper \cite{susy4}.
\section{Degeneracy of the spectrum}

In the previous sections, we have succeeded to construct 
the $N=2n$ supercharges and found the connection conditions 
compatible with supersymmetry.
In this section, we study the degeneracy of the spectrum, 
in particular, vacuum states with vanishing energy.

We first note that ${\cal G}_{{\cal R}_k}$ $(k=1,2,\cdots,n)$ 
commute with $H$ and also with each other.
These facts guarantee that we can introduce simultaneous 
eigenfunctions of the Hamiltonian and ${\cal G}_{{\cal R}_k}$ 
$(k=1,2,\cdots,n)$ such that
\begin{eqnarray}
	H \varphi_{E;\lambda_1,\cdots,\lambda_n}(x) &=& 
	  E \varphi_{E;\lambda_1,\cdots,\lambda_n}(x), \\
	{\cal G}_{{\cal R}_k} \varphi_{E;\lambda_1,\cdots,
	  \lambda_n}(x) &=&
	  \lambda_k \varphi_{E;\lambda_1,\cdots,\lambda_n}(x)
\end{eqnarray}
with $\lambda_k=1$ or $-1$ for $k=1,2,\cdots,n$.
Since $Q_a$ $(a=1,2,\cdots,2n)$ and ${\cal G}_{{\cal R}_k}$ 
$(k=1,2,\cdots,n)$ satisfy the relations
\begin{eqnarray}
	Q_a {\cal G}_{{\cal R}_k}=
		\left\{
			\begin{array}{lll}
				-{\cal G}_{{\cal R}_k}Q_a & \quad {\rm if} &
				\ a=2k-1 \ {\rm or}  \ 2k, \\
				+{\cal G}_{{\cal R}_k} Q_a & \quad {\rm if} & 
				\ {\rm otherwise},
			\end{array}
		\right.
\end{eqnarray}
the states $Q_{2k-1}\varphi_{E;\lambda_1,\cdots,\lambda_n}(x)$ 
and $Q_{2k} \varphi_{E;\lambda_1,\cdots,\lambda_n}(x)$ should 
be proportional to \\$\varphi_{E;\lambda_1,\cdots,-\lambda_k,
\cdots,\lambda_n}(x)$ , i.e.,
\begin{eqnarray}
	Q_{2k-1}\varphi_{E;\lambda_1,\cdots,\lambda_n}(x)
	=  -i \lambda_k Q_{2k}\varphi_{E;\lambda_1,\cdots,\lambda_n}(x) 
		 \propto  \varphi_{E;\lambda_1,\cdots,-\lambda_k,\cdots,
	    	\lambda_n}(x) ,
\end{eqnarray}
when $E\ne 0$. 
This implies that the degeneracy of the spectrum for $E \ne 0$ 
is given by $2^n$.
This result can be obtained from an algebraic point of view; 
for a fixed nonzero energy $E$, $Q_a /\sqrt{E}$ 
for $a=1,2,\cdots,2n$ form the Clifford algebra, 
and the representation is known as $2^n$.

The above argument cannot apply for states with $E=0$.
This is because any state $\varphi_0 (x)$ with vanishing energy satisfies
\begin{equation}
	Q_a \varphi_0(x)=0 \quad {\rm for} \ ^{\forall}a=1,2,\cdots,2n.
\end{equation}
It is easy to show that there are $2^n$ formal solutions 
to the above equations
\begin{equation}
	\varphi_{0;\lambda_1,\cdots,\lambda_n}(x)=
	N_{\lambda_1 \cdots \lambda_n}
		\left[
			\prod_{k=1}^n \frac{1}{2} \left(
				1+\lambda_k {\cal G}_{{\cal R}_k}\right) 
		\right]
		e^{-\lambda_1 \cdots \lambda_n W(x)}
		\label{v-state}
\end{equation}
with $\lambda_k=1$ or $-1$ for $k=1,2,\cdots,n$.
Here, $N_{\lambda_1 \cdots \lambda_n}$ denote normalization constants.
For a noncompact space, any non-normalizable states would be removed 
from the Hilbert space.
The space is, however, compact (a circle) in our model, so that 
the solutions (\ref{v-state}) are always normalizable.
Nevertheless, some of them must be removed from the Hilbert space.
This occurs due to incompatibility with the connection 
conditions (\ref{typeI}) and (\ref{typeII}).
 
The zero energy states for the type I connection conditions 
with $U_{\rm I} (+) \left( U_{\rm I}(-)\right)$ are given 
by $\varphi_{0;\lambda,\cdots,\lambda_n}(x)$ with $\lambda_1 
\lambda_2 \cdots \lambda_n=+1$ $(-1)$.
The remaining states with $\lambda_1\lambda_2
\cdots \lambda_n=-1$ $(+1)$ 
do not satisfy the connection conditions, and hence they must 
be removed from the Hilbert space.
Thus, the zero energy vacua are $2^{n-1}$-fold degenerate, 
and supersymmetry is unbroken\footnote{By analogy with 
supersymmetric quantum field theory, we say that supersymmetry 
is spontaneously broken if the action of the supercharges 
on any vacuum is nonvanishing.}.
For the type II connection conditions, all the states 
(\ref{v-state}) are found to be inconsistent with the 
connection conditions, so that there are no vacuum states 
with zero energy.
There is, however, an exception.
If the following relations are satisfied
\begin{eqnarray}
	\sqrt{\frac{1-a_3}{1+a_3}} &=& e^{W(l_0)-W(l_1)} ,\nonumber \\
	a_1&=& \sqrt{1-(a_3)^2}, \nonumber \\
	a_2 &=&0,
\end{eqnarray}
all the states (\ref{v-state}) accidentally become supersymmetric 
vacuum states compatible with the connection conditions. 
Therefore, for the type II connection conditions, 
supersymmetry is spontaneously broken except 
for the above case.
\section{Summary and discussions}
In this Letter, we have constructed the $N=2n$ supercharges 
and found a class of the connection conditions compatible 
with supersymmetry in one-dimensional quantum mechanics 
on a circle with $2^n$ point singularities.
The supercharges are represented in terms of the discrete 
transformations $\{ {\cal P}_k, {\cal Q}_k, {\cal R}_k \}$ 
$(k=1,2,\cdots,n)$. 
The action of  $\{ {\cal P}_k, {\cal Q}_k, {\cal R}_k \}$, 
in general, makes wavefunctions discontinuous, 
so that our realization of the $N=2n$ supersymmetry reflects 
the characteristics of singularities in quantum mechanics.

In our analysis, we required that all the $2n$ supercharges 
are 
hermitian and well-defined on the Hilbert space.
We can, instead, require that only a subset of them are 
hermitian and well-defined to reduce the $N=2n$ supersymmetry.
In other words, we allow some of the $2n$ supercharges to become 
ill defined due to connection conditions. 
This implies that the introduction of a number of point 
singularities can lead to a wide variety of $N$-extended 
supersymmetric models for any integer $N$.

It is interesting to notice that there exists one more 
discrete transformation ${\cal P}_{n+1}$ that produces 
singularities at $x=l_s$ for $s=0,1,\cdots,2^{n}-1$ but 
no other points.
Adding ${\cal P}_{n+1}$ to the algebra, we can construct 
$2n+1$ supercharges that form an $N=2n+1$ superalgebra \cite{susy4}.
Any subset of the $2n+1$ supercharges does not, however, 
coincide with the $N=2n$ supercharges in Eqs. (\ref{2ns-charges}) 
(unless the Hamiltonian is free).
Thus, the $N=2n+1$ supersymmetry including ${\cal P}_{n+1}$ 
in the algebra belongs to a different class from the $N=2n$ 
supersymmetry considered in this Letter. 
Full details 
will be discussed in a forthcoming paper \cite{susy4}.

\section*{Acknowledgements}
The authors wish to thank
C.S. Lim,
S. Tanimura,
I. Tsutsui
for valuable discussions and useful comments. 
This work was supported in part by the Grant-in-Aid for
Scientific Research (No.15540277) by the Japanese Ministry
of Education, Science, Sports and Culture.
K.T. is supported by The 21st century COE program named 
\lq\lq Towards a new basic science : depth and synthesis".

\baselineskip 5mm 

\pagebreak
\begin{figure}
\begin{center}
	\includegraphics{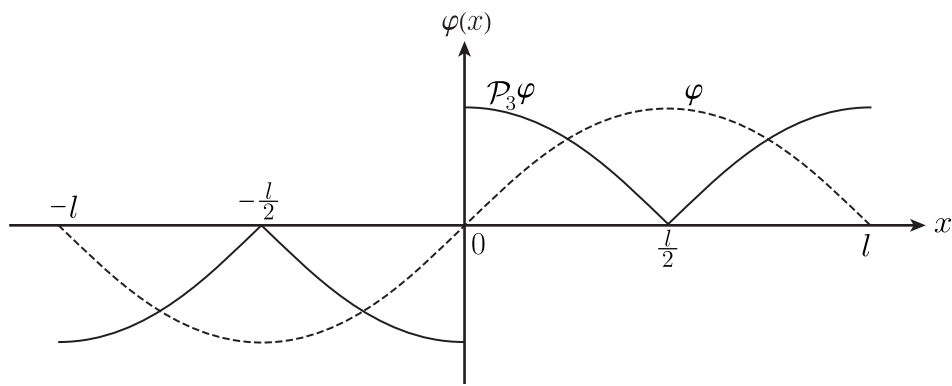}
\end{center}
\caption{ An example of the action of ${\cal P}_3$ on a function 
$\varphi(x)$. The dashed line denotes the original function 
$\varphi(x)$, and the solid line denotes $({\cal P}_3 \varphi )(x)$.}
\end{figure}
%
%
\begin{figure}
\begin{center}
	\includegraphics{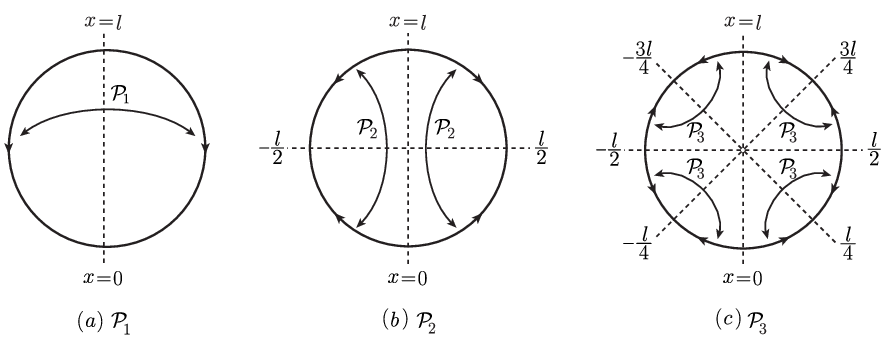}
\end{center}
\caption{ The geometrical meanings of ${\cal P}_k$ for $k=1,2$ and 3.}
\end{figure}
%
%
\begin{figure}
\begin{center}
	\includegraphics{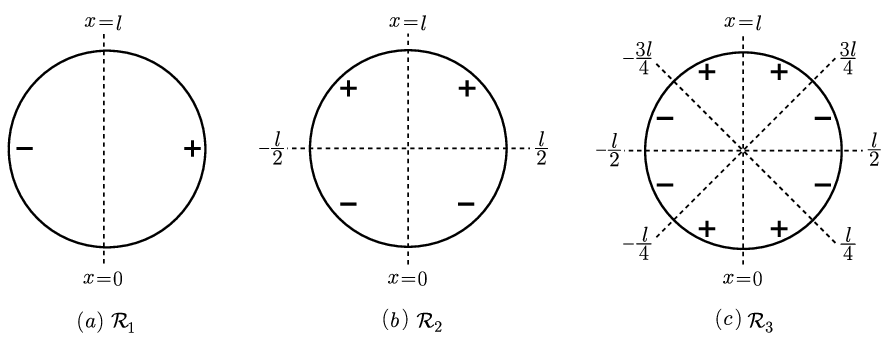}
\end{center}
\caption{ The geometrical meanings of ${\cal R}_k$ for $k=1,2$, and 3.}
\end{figure}


\begin{thebibliography}{99}
\bibitem{u(2)1}
    M. Reed, B. Simon,
    {\sl \lq\lq Methods of Modern Mathematical Physics"},
    Vol.II, Academic Press, New York, 1980.
\bibitem{u(2)2}
    P. $\check{S}$eba, Czeck. J. Phys. {\bf 36} (1986) 667.
\bibitem{u(2)3}
    S. Albeverio, F. Gesztesy, R. H{\o}egh-Krohn, H. Holden,
    {\sl \lq\lq Solvable Models in Quantum Mechanics"},
    Springer, New York, 1988.
\bibitem{duality1}
    T. Cheon, T. Shigehara, Phys. Rev. Lett. {\bf 82} (1999) 2536,
    quant-ph/9806041.
\bibitem{duality2}
    I. Tsutsui, T. F{\"u}l{\"o}p, T. Cheon, J. Phys. Soc. Jap.
    {\bf 69} (2000) 3473, quant-ph/0003069.
\bibitem{berry1}
    T. Cheon, Phys. Lett. {\bf A 248} (1998) 285, quant-ph/9803020.
\bibitem{berry2}
    P. Exner, H. Grosse,
    {\sl \lq\lq Some properties of the one-dimensional generalized point
interactions (a torso)},
    math-ph/9910029.
\bibitem{anomaly}
 T. Cheon, T. F{\"u}l{\"o}p, I. Tsutsui, Ann. Phys. {\bf 294}
   (2001) 1, quant-ph/0008123.
\bibitem{susy1}
    T. Uchino, I. Tsutsui,
	Nucl. Phys. {\bf B 662} (2003) 447,
    quant-ph/0210084.
\bibitem{susy2}
    T. Nagasawa, M. Sakamoto, K.Takenaga,
	Phys. Lett. {\bf B 562} (2003) 358,
    hep-th/0212192.
\bibitem{susy3}
    T. Uchino, I. Tsutsui,
	J. Phys. {\bf A 36} (2003) 6493,
    hep-th/0302089.
\bibitem{susy5}
    T. F{\"u}l{\"o}p, I. Tsutsui,  T. Cheon, 
	{\sl \lq\lq Spectral Properties on a Circle
	with a Singularity"},  
	quant-ph/0307002.
\bibitem{susy4}
    T. Nagasawa, M. Sakamoto, K. Takenaga, in preparation.
	

\end{thebibliography}
\end{document}